\documentclass{aastex631}
\usepackage{amsmath}

\received{March, 2022}
\revised{February, 2023}
\submitjournal{AJ}

\shorttitle{BASS Astrometry}
\shortauthors{PENG et al.}

\begin{document}

\title{Astrometric Calibration of the Beijing-Arizona Sky Survey}

\correspondingauthor{Xiyan Peng, Zhaoxiang Qi, Zhenyu Wu}
\email{xypeng@shao.ac.cn, kevin@shao.ac.cn, zywu@nao.cas.cn}
\author{Xiyan Peng }
\affiliation{Shanghai Astronomical Observatory, \\
 Chinese Academy of Sciences, \\
  Shanghai 200030, China \\}

\author{Zhaoxiang Qi}
\affiliation{Shanghai Astronomical Observatory, \\
 Chinese Academy of Sciences, \\
  Shanghai 200030, China \\}
\affiliation{School of Astronomy and Space Science,\\
University of Chinese Academy of Sciences, \\
 Beijing 100049, China\\}
 
\author{Tianmeng Zhang}
\affiliation{National Astronomical Observatories, \\
  Chinese Academy of Sciences, Beijing 100012, China\\}
\affiliation{School of Astronomy and Space Science,\\
University of Chinese Academy of Sciences, \\
 Beijing 100049,  China\\}
 
 \author{Zhenyu Wu }
\affiliation{National Astronomical Observatories, \\
  Chinese Academy of Sciences, Beijing 100012, China\\}
  \affiliation{School of Astronomy and Space Science,\\
University of Chinese Academy of Sciences, \\
 Beijing 100049,  China\\}

 \author{Zhimin Zhou}
\affiliation{National Astronomical Observatories, \\
  Chinese Academy of Sciences, Beijing 100012, China\\}
 
 \author{Jundan Nie}
\affiliation{National Astronomical Observatories, \\
  Chinese Academy of Sciences, Beijing 100012, China\\}

 \author{Hu Zou}
\affiliation{National Astronomical Observatories, \\
  Chinese Academy of Sciences, Beijing 100012, China\\}
\author{Xiaohui Fan}
\affiliation{Steward Observatory, University of Arizona, \\
Tucson, AZ 85721, USA\\} 
\author{Linhua Jiang}
\affiliation{Kavli Institute for Astronomy and Astrophysics, Peking University,\\
 Beijing 100871, China\\}
\author{Ian McGreer}
\affiliation{Steward Observatory, University of Arizona, \\
Tucson, AZ 85721, USA\\} 
\author{Jinyi Yang}
\affiliation{Steward Observatory, University of Arizona, \\
Tucson, AZ 85721, USA\\}
\author{Arjun Dey}
\affiliation{National Optical Astronomy Observatory, \\
Tucson, AZ 85719, USA}
\author{Jun Ma}
\affiliation{National Astronomical Observatories, \\
  Chinese Academy of Sciences, Beijing 100012, China\\}
\author{Jiali Wang}
\affiliation{National Astronomical Observatories, \\
  Chinese Academy of Sciences, Beijing 100012, China\\}
\author{David Schlegel}
\affiliation{Lawrence Berkeley National Labortatory, \\
Berkeley, CA 94720, USA}
\author{Xu Zhou}
\affiliation{National Astronomical Observatories, \\
  Chinese Academy of Sciences, Beijing 100012, China\\}

\nocollaboration{16}

\begin{abstract}

We present the astrometric calibration of the Beijing-Arizona Sky Survey (BASS). The BASS astrometry was tied to the International Celestial Reference Frame via the \emph{Gaia} Data Release 2 reference catalog.  For effects that were stable throughout the BASS observations,  including differential chromatic refraction and the low charge transfer efficiency of the CCD, we corrected for these effects at the  raw image coordinates. Fourth-order polynomial intermediate longitudinal and latitudinal corrections were used to remove optical distortions. The comparison with the \emph{Gaia} catalog shows that  the systematic errors, depending on color or magnitude, are less than 2 milliarcseconds (mas).   The position systematic error is estimated to be about $-0.01\pm0.7$ mas in the region between 30 and 60 degrees of declination and up to $-0.07 \pm 0.9$ mas in the region north of declination 60 degrees.

\end{abstract}

\section{Introduction} \label{sec:intro}

The Beijing-Arizona Sky Survey (BASS) is a photometric survey of the northern Galactic Cap using the 90Prime imager mounted on the 2.3 m Bok telescope at Kitt Peak. It is a four-year collaboration between the National Astronomical Observatories, Chinese Academy of Sciences, and Steward Observatory, University of Arizona, for the Dark Energy Spectroscopic Instrument project target selection (DESI, \citealt{desi2016}). The BASS began an imaging survey in January 2015. The sky area coverage was approximately 5400$^2$, which was scanned through three passes using the Sloan Digital Sky Survey (SDSS, \citealt{2000AJ....120..1579V}) \textit{g} filter and a new \textit{r} filter  close to that used in the Dark Energy Survey \citep{desi2005}.  Combining with BASS survey and the other surveys, we can study Galactic structures, galaxy clusters, AGN evolution, high redshift quasars, and large-scale structure of the universe.

Astrometric calibration is the cornerstone step in the reduction of large-scale image surveys. To recover most of the photometric quality of the single-epoch images, BASS performed forced photometry on the single-epoch exposure using the positions determined in the stacked images. The success of this approach requires precise astrometry. Searches for transient sources also take advantage of precise astrometry to improve the subtraction of the static sources. Given a more than 10-year time baseline between the imaging data from SDSS \citep{2000AJ....120..1579V} and BASS surveys, proper motions measured from SDSS and BASS could reach accuracies of a few mas per year for stars two mag fainter than the \emph{Gaia} limits. The motivation and practice of astrometry using large-format ground-based CCD cameras have been discussed by \citet{2013A&A...554...101},  \citet{2016magnier}, and \citet{2017PASP...129...4503B}.   

During the BASS data processing,  we find some effects that affect the accuracy of astrometry. First, the geometric distortion of the BASS that changes over time is a problem because the distortion solution limits the accuracy of the final astrometry. In addition, from Figure 10 in \citet{2017AJ.153..276V}, there is a clear discontinuity in the middle of  the astrometric residual maps in the declination (decl.), while there is no similar apparent substructures in the right ascension (R.A.). This is because the Bok telescope CCDs have a relatively low charge transfer efficiency (CTE), leading to the centroids of the sources gradually deviating from the initial positions as transfer occurs. Finally, Differential Chromatic Refraction (DCR) affects the BASS astrometry because for a given star of a color different from the reference stars, as exposures are taken at a higher air mass, the apparent position of the star will be biased along the parallactic angle \citep{eg15}. In this work, we will show how to calculate and remove the influence of these effects.

The remainder of this paper is organized as follows. In Section 2, we describe the data used in BASS astrometry. In Section 3, we describe how to calculate the BASS distortion map and measure the low CTE and DCR effects by analyzing the residual error between BASS and \emph{Gaia}. Section 3 also shows how to remove the system effects in BASS astrometry. Section 4 shows the extent to which system influence can be eliminated. The accuracy of the BASS astrometry is also presented in Section 4. Finally, Section 5 concludes the study.

\section{DATA} \label{sec:style}
\subsection{The instrument and observation}
\begin{figure*}
\centering
\includegraphics[width=10cm]{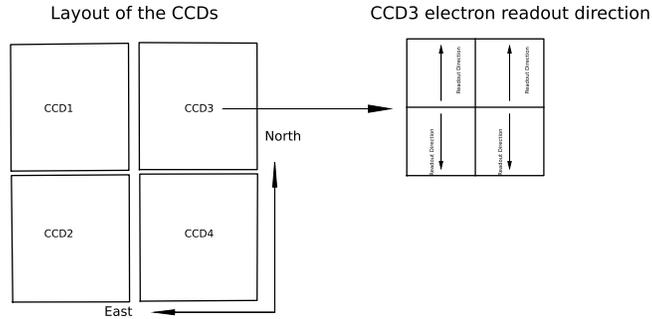}
\caption{The left panel shows the layout of the 90Prime CCD mosaic imaging plane. The labels represent the CCD numbers. Each CCD has four readout amplifiers and two different readout directions. As an example, the right panel shows the electron readout direction for CCD3. The image is read out by amplifiers along the direction of the arrows.}
\label{ccdaread}
\end{figure*}

The BASS is carried out by the Bok 90 inch telescope at the Steward Observatory using its 90Prime wide-field imager \citep{2018ApJS....237..2V}. As shown in the left panel of Figure \ref{ccdaread}, the imaging camera consists of four 4096 $\times$ 4032 pixel CCDs with a pixel scale of approximately $0.454''$. Each BASS image covers $1.08 \times 1.03$ deg$^2$. The gap between the CCDs is roughly $170''$ in R.A. and $55''$ in decl.   Each CCD was read out by four amplifiers along two different directions, with a readout time of 17 s for an unbinned full image at a pixel rate of 250 kHz per amplifier. As an example, the electron readout direction for CCD3 is shown in the right panel of Figure \ref{ccdaread}. After the observation, high-quality single-epoch images were saved to ensure that the three-pass co-added imaging depths met the requirements. Table 1 presents an overview of the Bok 90 inch telescope, its camera, and the data acquisition.

\begin{deluxetable*}{cc}
\tablewidth{0pt}
\tablecaption{Overview of the telescope, camera and observation} 
\tablehead{
\colhead{Parameter} & \colhead{Value} \\
}
\colnumbers
\startdata
Primary mirror clear aperture        &    229  mm                       \\
Primary focal ratio                          &          f/2.66                            \\
Field of view                                    &            65$'$ $\times$ 62$'$    \\
CCD number                                   &            4                                    \\
Number of pixels                             &            4096  $\times$  4032      pixels \\
Plate scale                                      &            30.2 $''$/mm  or 0.454$''$ /pixel   \\
Frames per field                              &            Two bands $\times$ three times  \\
Wavelengths  of $g$                 &            4776 $\AA$                      \\
Wavelength  of $r$                &            6412 $\AA$                      \\
Expected depth (5$\sigma$ 3 exposures)            &    $g$ = 24.1, $r$ = 23.5             \\
Typical seeing                                &            1.7$''$         
\enddata
\end{deluxetable*}

\subsection{Image Reduction}
The image-processing steps are described in detail in the BASS Data Release 1 (DR1, \citealt{2017AJ.153..276V}),  the BASS Data Release 2 (DR2, \citealt{2018ApJS....237..2V})  and  the BASS Data Release 3 ( DR3, \citealt{2019ApJS....245..4V}). Crosstalk, bias, and flat calibrations were applied to the individual chips. All BASS image data products were stored in multi-extension FITS files, with each image associated with a pixel-level weight map and mask map. Weight and mask maps were generated by image reduction and carried forward at each stage of data processing. The weight map describes the noise intensity at each pixel in the astronomical images. Bad pixels, cosmic rays, satellite trails, black cores, and bleeding trails of saturated star masking were carried out and recorded in a mask image. Mask maps use integer numbers to flag bad and noisy pixels.    

\subsection{ Object Detection and Characterization}
For the BASS astrometry, the input data detected from the image contains the measured position $x$, $y$, the instrumental magnitude, the calibrated magnitude of objects, and the corresponding error information.
SExtractor \citep{1996A&AS...117...393B} was used to detect and characterize  stars and other objects in the BASS images. Detection parameters (e.g., threshold, characteristic object size, and ultimately the signal-to-noise ratio) are set to find bright sources while reducing the number of spurious sources. Weight and mask maps were used in the progress of SExtractor detection. For CCD images, it has been verified that for isolated, the Gaussian-like point-spread function (PSF), which describes the precision in centroiding offered by the windowed centroids (XWIN$\_$IMAGE, YWIN$\_$IMAGE), approaches the accuracy of ideal PSF-fitting astrometry \citep{1996A&AS...117...393B,2017PASP...129...4503B}. All the inputs $x$ and $y$ measured are XWIN$\_$IMAGE and YWIN$\_$IMAGE from the SExtractor. The normalized instrumental magnitude, $M\rm{_{inst}}$ = $M\rm{_{auto}}$ + 25.5, was stored for each measurement. $M\rm{_{auto}}$ is the automatic aperture magnitude from the SExtractor. The value of 25.5 is an arbitrary (but fixed) constant offset used to place the instrumental magnitudes within the approximately correct range of the $g$ and $r$ bands. The instrumental magnitude was used for the calculation and correction of the low CTE. The photometric calibration of the BASS is presented in the work of  \cite{2018PAP....130..850V}. The calibrated magnitude was used for the calculation and correction of the DCR.

\subsection{Reference catalog}

Several reference catalogs were used during the survey. For BASS DR1, the reference catalogs were SDSS and the point-source catalog of the Two Micron All Sky Survey (2MASS; \citealt{2006AJ..131...1163}). The SDSS catalog contains more faint sources than other catalogs and covers most of the BASS area. Therefore, the SDSS catalog was used as the main reference catalog. If a region was outside the SDSS coverage, 2MASS was used as the reference catalog. Since April 25, 2018, the \emph{Gaia} data release 2 catalog has been available \citep{Lindegren18}. Because of the high accuracy  and high density, the \emph{Gaia} catalog is an excellent tool for detecting systematic errors in the BASS data. From the BASS DR2, the \emph{Gaia} catalog is used as the reference catalog. \emph{Gaia} proper motions have been used to propagate the \emph{Gaia} positions to the epoch of the individual BASS observations.

\section{The BASS Astrometry}

\begin{figure*}
\centering
\includegraphics[width=8cm]{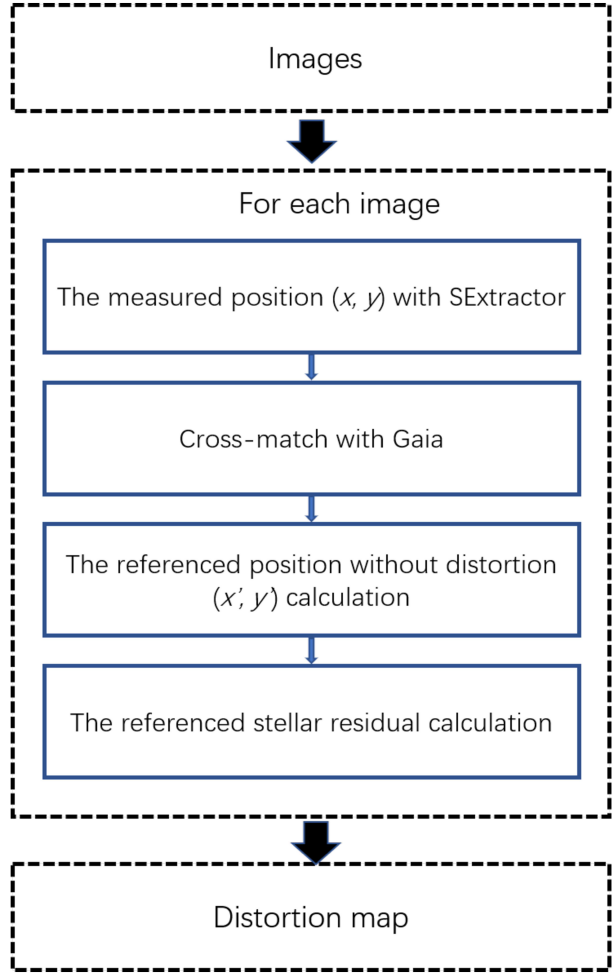}
\caption{The flowchart summarizes the specific steps used to derive geometric distortion.  }
 \label{flow1}
\end{figure*}

\begin{figure*}
\centering
\includegraphics[width=18cm]{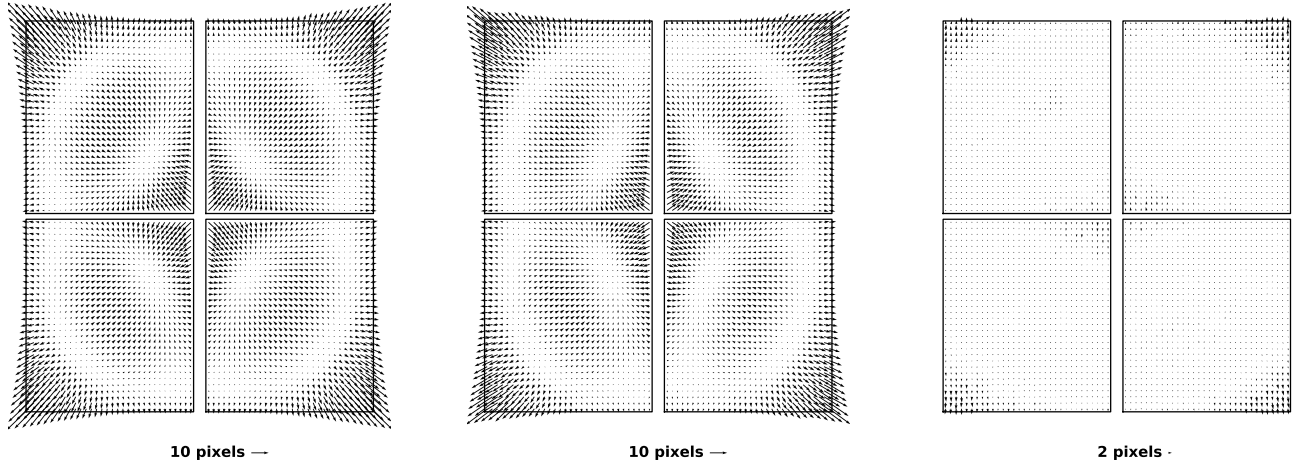}
\caption{Maps of the geometric distortion for BASS observations over a night.  The left panel shows the distortions map from the first half of the night exposures, and the middle panel shows the distortions map from the second half of the night exposures, and the right panel shows the difference between the two maps. The geometric distortion varies in one night up to 2 pixels $\sim 1 ''$. }
 \label{mu}
\end{figure*}

During the BASS DR1 and the BASS DR2, we found some system errors in the BASS astrometry. In this work, we studied these system errors, including  low CTE, DCR, and optical distortions, and corrected them. Sections 3.1 - 3.3 focus on the calculation of various effects.  Section 3.4 shows how to correct for the various effects. 

\subsection{ Field Distortion}

We make use of the \emph{Gaia} catalog to derive the BASS geometric distortion. Residuals between the measured positions ($x, y$) from the BASS and the reference  measured positions  ($x'$, $y'$)  from \emph{Gaia} without distortion correction  were used to obtain geometric distortion.  The reference measured positions ($x'$, $y'$) without distortion correction are calculated based on the assumption that without distortion, the coordinate transformation between the intermediate standard positions of the reference stars and measured positions is linear. Similar methods are available in \cite{2003PASP...115...113}, \cite{2006AA...454...1029}, \cite {2012AJ..144..170}, and \cite{2019MNRAS.485.1626}.  To derive the geometric distortion, we average the stochastic atmospheric signal by stacking and binning the residuals from multiple exposures. Then, the field distortion map was determined by averaging the residuals within 1024 individual bins (32, 32 across the $x$, $y$ field). \emph{Gaia} stars with a magnitude range from 16 to 21 are used as reference stars.  Figure \ref{flow1} summarizes the specific steps used to derive the geometric distortion.   

Figure \ref{mu}  shows the $r$-band field distortion maps for the Bok telescope over one night.  The left panel shows the distortion map from the first half of night exposure, and the middle panel shows the distortion map from the second half of night exposure. As expected from the optical design, the distortion pattern was radial.  In addition, these  panels reveal clear large-scale systematic errors across the focal plane, with residuals of up to 37 pixels in the corner.  The right panel shows the difference between the two maps.  The geometric distortion varies by up to 2 pixels in one night, which is about 1$''$. In addition, from the distortion map of various subsets of the data, we found that the distortions are unstable during the BASS observation, depending on the observation time and filter. The possible cause of the instability distortion may be a change in focus or  telescope flexure. 

For each BASS image, geometric distortion correction was performed after CTE and DCR correction. Because the distortion is constantly changing, the polynomial used to correct the distortion changes with the image.  For each BASS image, we used SCAMP (Bertin et al. 2006) along with the reference catalog for astrometric calibration, and assigned each CCD a fourth-order polynomial distortion function around the tangent plane for R.A. and decl. after CTE and DCR correction, see Section 3.4  and Appendix A for details. We do not fit using higher order polynomials because these sometimes resulted in overfitting especially in sparse fields.

\subsection{Low CTE}

As the CCD reads an image, electrons are transferred from pixel to pixel until the charge reaches the output register.  When transfer occurs, a low CTE causes electrons to be left behind, leading to a slight centroid shift \citep{2006PASP....793..127V}.  The position, magnitude, and background of the sources were directly related to  the associated centroid shift caused by low CTE. For BASS, there is a significant CTE effect along the $y$-axis because of the large number of transfers, opposite readout direction, and low CTE.  High-quality BASS frames containing approximately several thousand stars observed in 2016 were selected as samples. Images with similar sky backgrounds suffer a similar CTE effect, and we study CTE loss separately for frames with different sky backgrounds.  The samples were sorted by sky background and divided into 10 subsamples of the same size. Table 2 lists the background intervals for the subsamples. $bk_{min}$ and  $bk_{max}$  are the minimum and maximum values of the sky background, respectively.

\begin{deluxetable*}{ccc}
\tablewidth{0pt}
\tablecaption{Sky background intervals } 
\tablehead{  
\colhead{Sample} & \colhead{The minimum value of background}  & \colhead{The maximum value of background}     \\
}
\colnumbers
\startdata
subsample 1        &      0 ADU      &   549 ADU                 \\
subsample 2        &     549 ADU   &   645 ADU                      \\
subsample 3        &     645 ADU   &   755 ADU  \\
subsample 4        &     755 ADU   &   914 ADU                                 \\
subsample 5        &     914 ADU   &   1095 ADU    \\
subsample 6        &     1095 ADU &  1342 ADU  \\
subsample 7        &     1342 ADU &  1680 ADU \\
subsample 8        &     1680 ADU &   2138 ADU                     \\
subsample 9        &      2138 ADU &   3218 ADU                      \\
subsample 10      &      3218 ADU &   40000 ADU          \\
\enddata
\end{deluxetable*}

\begin{figure*}
\centering
\includegraphics[width=10cm,angle=0]{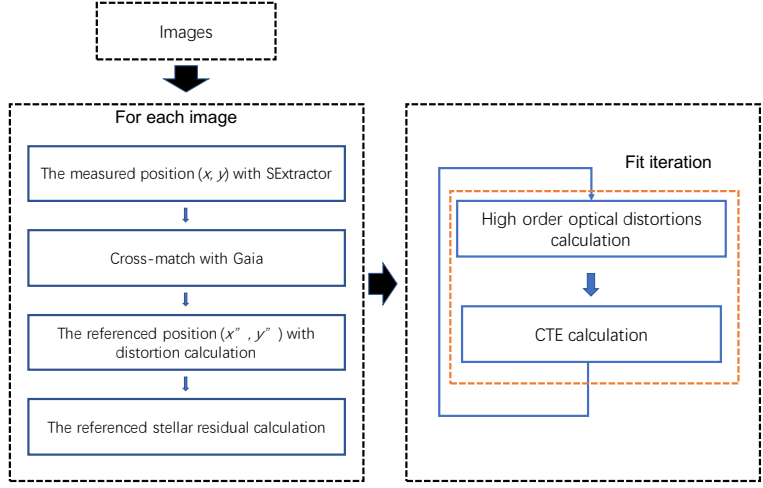}
\caption{The flowchart summarizes the specific steps used in CTE analysis.  }
 \label{flowcte}
\end{figure*}

Residuals between the measured positions ($ y$) from the BASS image and reference measured positions ($y''$) from Gaia are used to obtain the CTE effect. $y''$ are obtained using fourth-order polynomial distortion correction. The BASS discontinuous position shift because of the low CTE causes these residuals to contain additional high-order distortions. By ignoring the centroid shift because of DCR effects, we assume that these residuals include only CTE effects and high-order distortions. In Equation (\ref{2}), $\Delta y$ is the $y$ residual, $\rm{CTE}(mag,\it{y})$ is the centroid shift from the CTE effect, $f(y)$ is the high-order distortion effect. The CTE loss is parameterized according to the distance of the source from the output register and instrumental magnitude of the source using Equation (\ref{3}).  The CTE effects ($y <=2016$ and  $y >2016$) are fitted separately because the readout directions of the electrons are different. The high-order distortion on the $y$-axis was parameterized using Equation (\ref{4}). The CTE is solved iteratively for each CCD. After reapplying high-order distortion corrections and re-running the residuals, the corrections were continuously updated until the residuals were flattened out. Figure \ref{flowcte} summarizes the specific steps used in CTE analysis.

\begin{figure*}
\centering
\includegraphics[width=10cm]{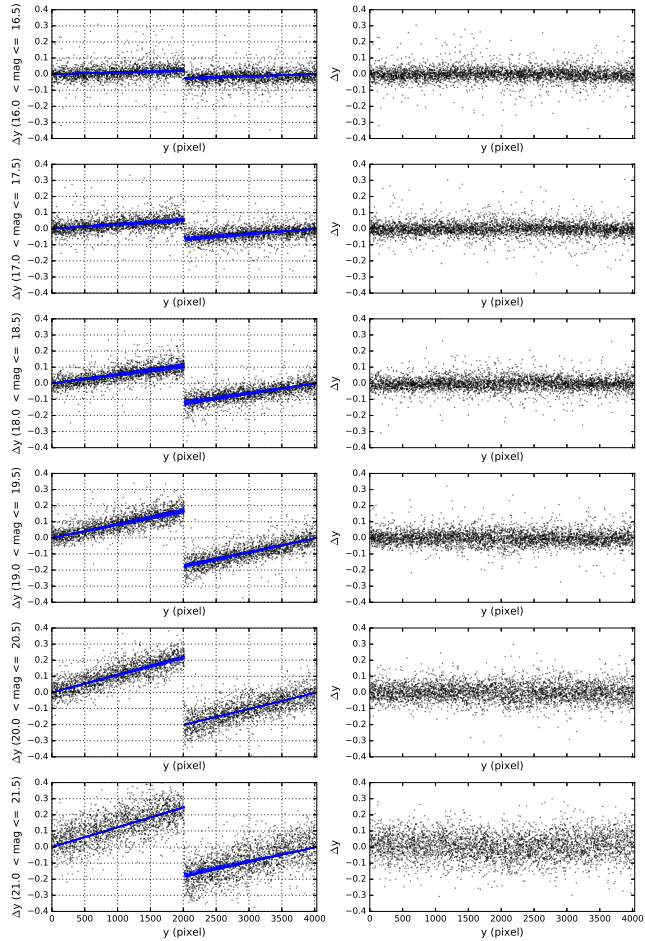}
\caption{Residuals in the $y$ direction before applying CTE correction (left panels) and the residuals after using CTE correction (right panels) are plotted against $y$ in the different magnitude range.  The blue dots in the lefts panel show the best fit for the CTE.}
\label{mucte1}
\end{figure*}

\begin{figure*}
\centering
\includegraphics[width=10cm]{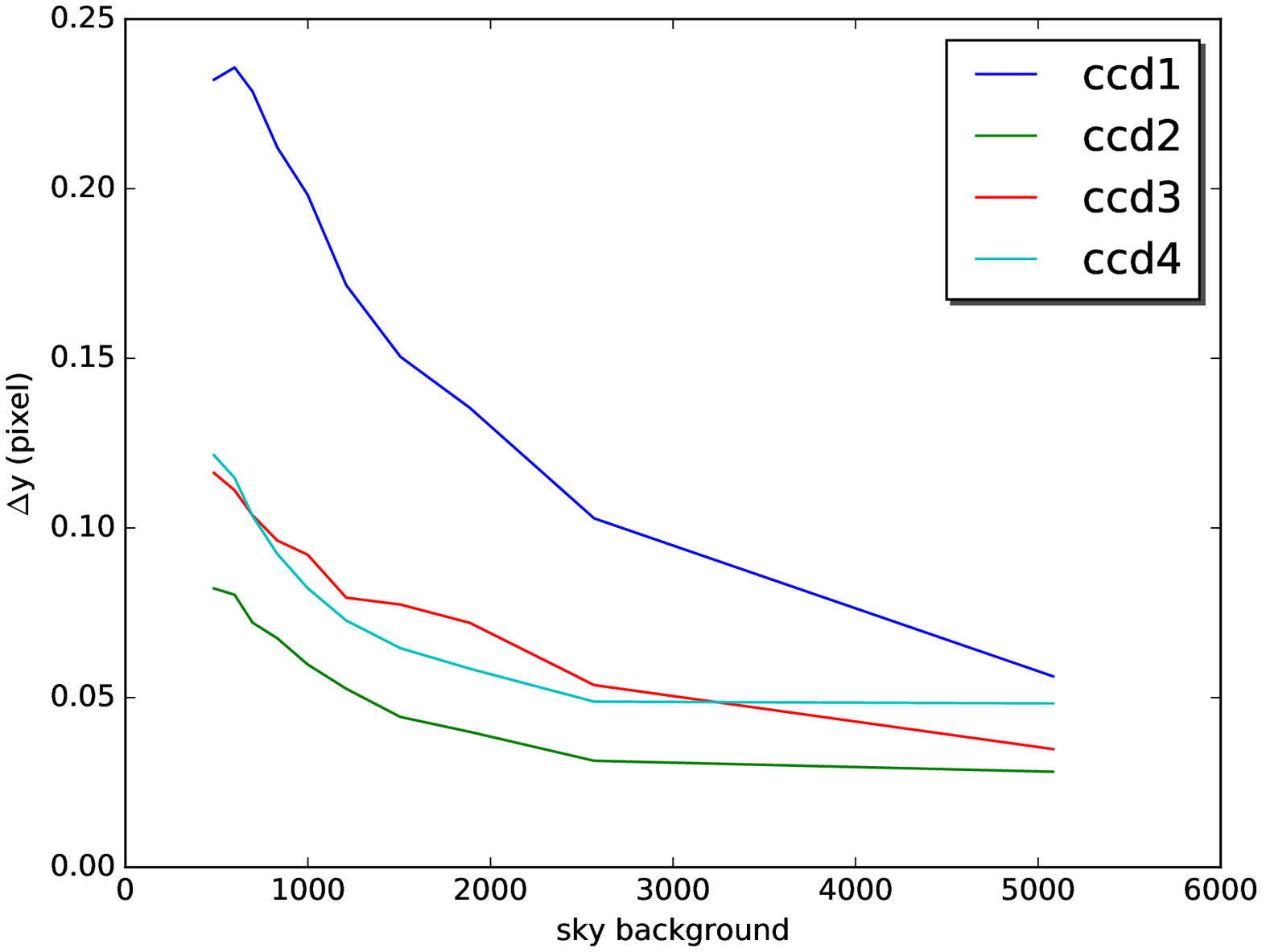}
\caption{CTE corrections for objects with magnitude = 20.6 and y=2000 as a function of sky background for different CCDs.}
\label{mubk}
\end{figure*}

All CCD detectors used in the 90Prime focus, particularly the CCD 1, have significant CTE effects along the $y$-axis, with a maximum of about 0.2 pixels. Figure \ref{mucte1} shows CTE loss in subsample 1 of CCD 1 as a function of star brightness and the $y$ position. The black dots in the left panel show the residuals in the $y$ position as a function of $y$ in the different instrumental magnitude ranges. The blue dots in the left panels indicate the fitted CTE effect. The right panels show the residuals at the $y$ position as a function of $y$ for different instrumental magnitude ranges after removing high-order optical distortions and the CTE effect. The figure clearly shows that CTE loss is associated with increased magnitude and location of the source on the CCD from the output register. With different sky backgrounds, the charge in each CCD suffers from different percentages of CTE loss. Figure $\ref{mubk}$ shows CTE correction for sources with mag=20.6 and position $y$=2000 as a function of the background. From Figure $\ref{mubk}$, we find that (1) the CTE effect decreases with increasing background, and (2) the CTE effect changes with the CCD.

\begin{eqnarray}
\Delta y& = &
\rm{CTE}(mag,\it{y}) +f(y) \label{2} 
\end{eqnarray}

\begin{eqnarray}
\rm{CTE}(mag,\it{y})  & = & 
(a +b \times \rm{mag}+c \times \rm{mag}^2+d \times \rm{mag}^3+e \times \rm{mag}^4) \times \it{y} 
  \text(y<=2016), \nonumber \\
\rm{CTE}(mag,\it{y})  & = & (f +g \times \rm{mag}+h \times mag^2+i \times mag^3+j \times mag^4) \times (4032-\it{y}) \text(y>2016)
 \label{3} 
\end{eqnarray}

\begin{eqnarray}
f(y) & = &
m_{0}+m_{1}y+m_{2}y^{2}+m_{3}y^{3}+m_{4}y^{4}
 \label{4} 
\end{eqnarray}

\subsection{DCR} 

The wavelength dependence of atmospheric refraction causes the elongation of finite-bandwidth images along the elevation vector. While images are taken from different bandpasses compared to the reference catalog, DCR will affect the measured positions of stars along the parallactic angle \citep{Hambly2001, Platais12,  Munn2014, 2016magnier}. For BASS astrometry, the effect of DCR on $g$-band images is greater than that on $r$-band images. For BASS $g$ band images, residuals between the measured positions from the BASS and the reference measured positions from $Gaia$ with CTE and distortion correction are used to obtain the DCR effect.

\begin{figure*}
\centering
\includegraphics[width=10cm]{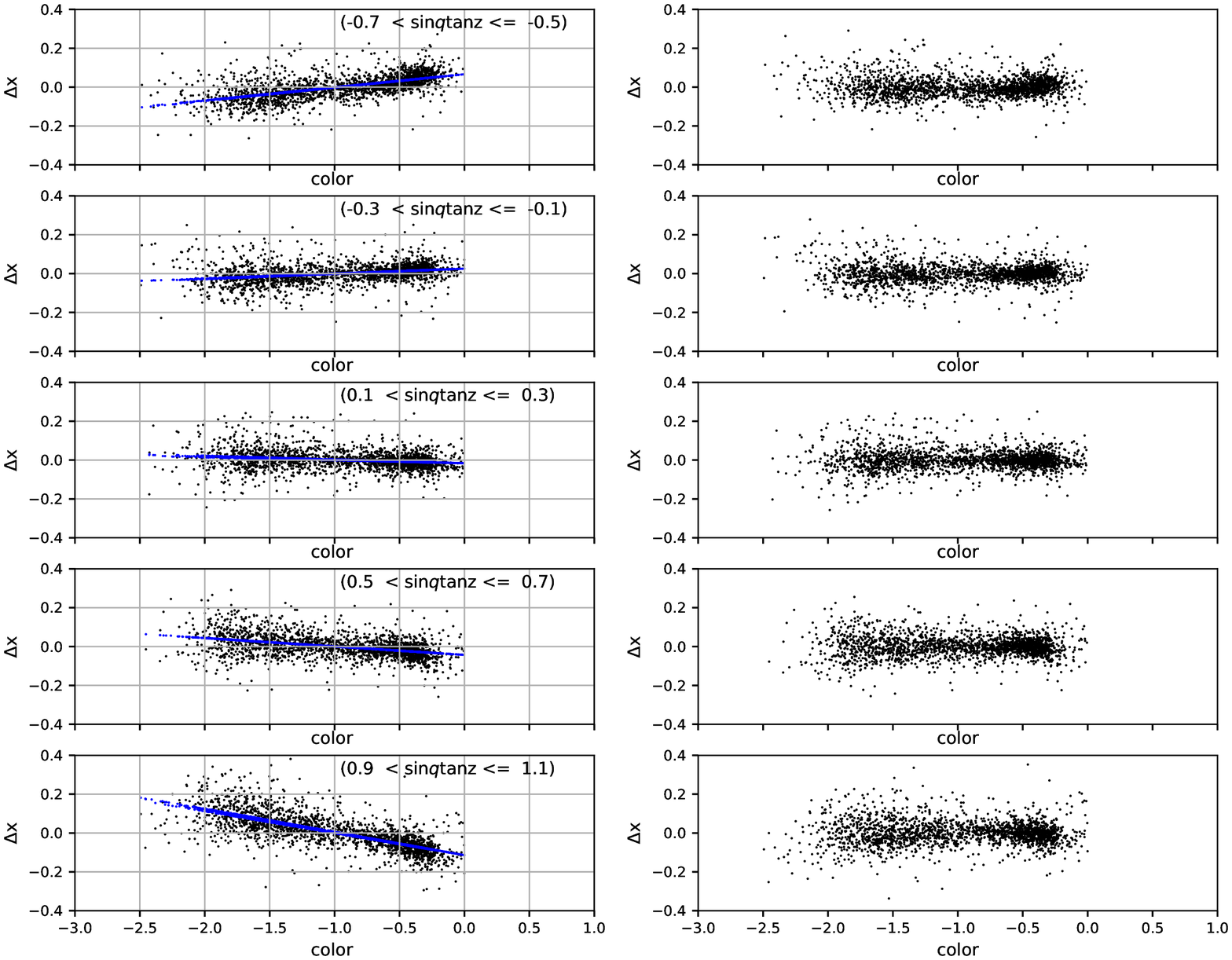}
\caption{Residuals before applying DCR correction (left panels) and the residuals after applying DCR correction (right panels) are plotted against color in different position range. The blue dots in left panels show the best fit to the DCR corrections. }
\label{nmu7}
\end{figure*}

The DCR correction was modeled as a function of color (\emph{Gaia} magnitude minus BASS calibrated magnitude), zenith distance, and parallactic angle using Equation (\ref{n10}). In Equation (\ref{n10}), $z$ is the zenith distance. $q$ is the parallactic angle.  $\rm{DCR}_{\it{x}}$ and $\rm{DCR}_{\it{y}}$ are the DCR corrections in the $x$ and $y$ directions, respectively. Figure \ref{nmu7} shows the effect of DCR in the $x$ direction as a function of color in the different $\sin q \times \tan z$ ranges.
The black dots in the left panels show the residuals in the $x$ direction as a function of color in different ranges of $\sin q \times \tan z$. The blue dots in the left panels represent the fitted DCR. 
The right panel shows the residuals in the $x$-direction after the DCR effect is removed. With the increase of $ \sin q \times \tan z $, the DCR correction in the $x$ direction keeps changing. 

\begin{eqnarray}
\rm{DCR}_{\it{x}}   & = &
(c_0 +c_1 \times \sin q \times \tan z) \times \rm{color} +c_2  \nonumber \\
\rm{DCR}_{\it{y}} & = &
(d_0 +d_1 \times \cos q \times \tan z) \times \rm{color}  + d_2
\label{n10} 
\end{eqnarray}

\subsection{Processing Pipeline} 

\begin{figure*}
\centering
\includegraphics[width=10cm]{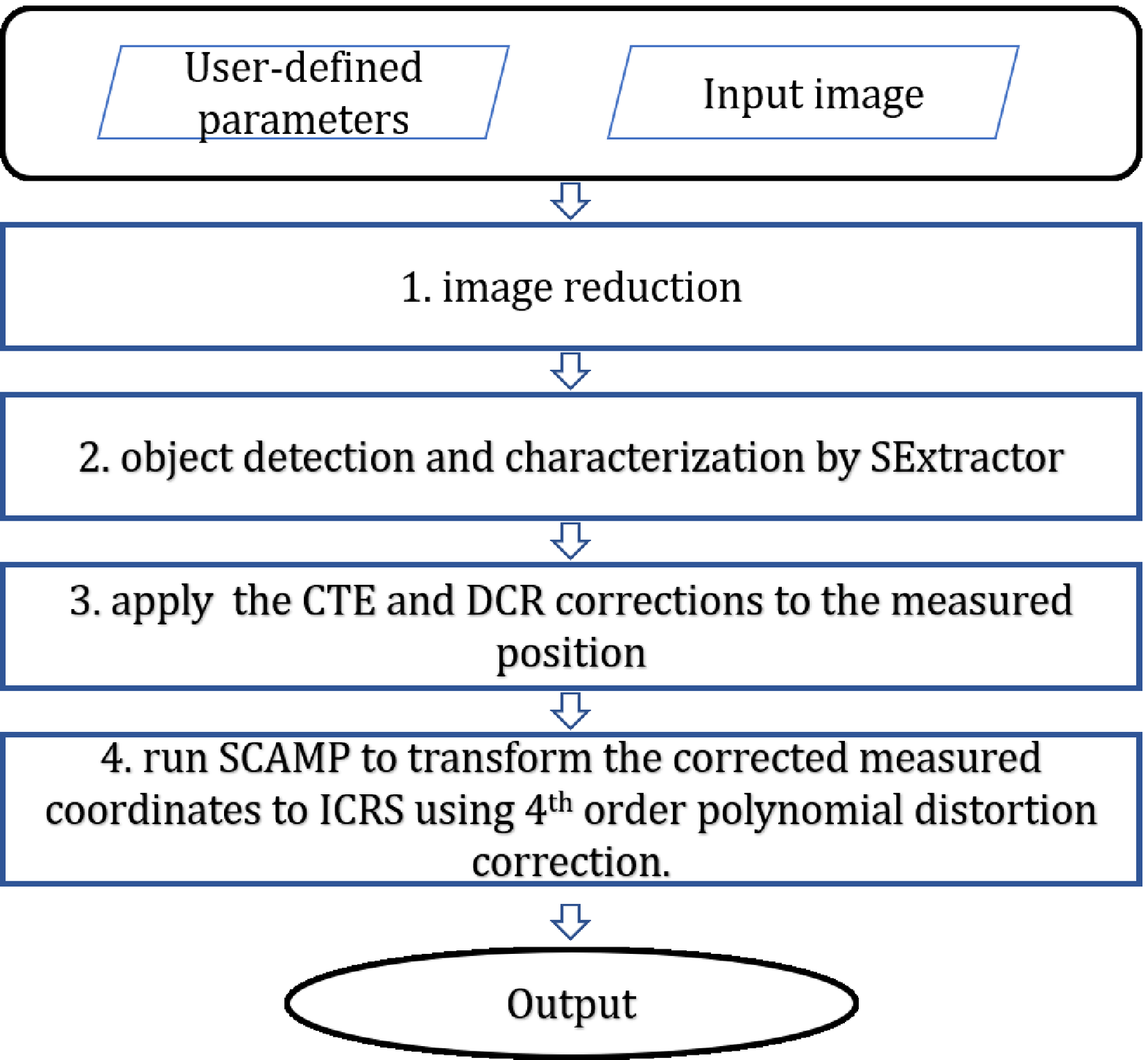}
\caption{The flowchart for BASS astrometry.}
\label{muflow}
\end{figure*}

We developed a pipeline based on the above studies to do BASS astrometry. Figure \ref{muflow} summarizes the specific steps used in the BASS astrometry pipeline. It primarily includes image reduction, object extraction and characterization, measurement coordinate correction, and astrometric calibration. The details of steps 1 and 2 are described in Sections 2.2 and 2.3. In step 3, the low CTE and DCR corrections were applied to the measured coordinates using Equation (\ref{n11}). $x$ and $y$ are the measured coordinates of the sources, and $x_c$ and $y_c$ are the corrected measured coordinates. In step 4, we use SCAMP along with the \emph{Gaia} catalog for the astrometric calibration. SCAMP reads the corrected measured coordinates, centroid errors, flux measurements, and flux error data and collects information including the starting World Coordinate System (WCS) information to compute astrometric projection parameters. We assigned each CCD a fourth-order polynomial distortion function around the tangent plane for both R.A. and decl. to calculate distortion. SCAMP calculates astrometric solutions with a TPV WCS, which is a non-standard convention following the WCS standard rules. It builds on standard TAN projection \citep{2002A&A... 395..1077}, by adding a general polynomial distortion correction, \emph{see Appendix A for details.} For each BASS exposure, SCAMP runs iteratively based on the last result to improve accuracy.

\begin{eqnarray}
x_c & = &
  x -\rm{DCR}_{\it{x}}    \nonumber \\
y_c & = &
 y - \rm{DCR}_{\it{y}}  - CTE(mag,y)
\label{n11} 
\end{eqnarray}

\begin{figure*}
\centering
\includegraphics[width=10cm]{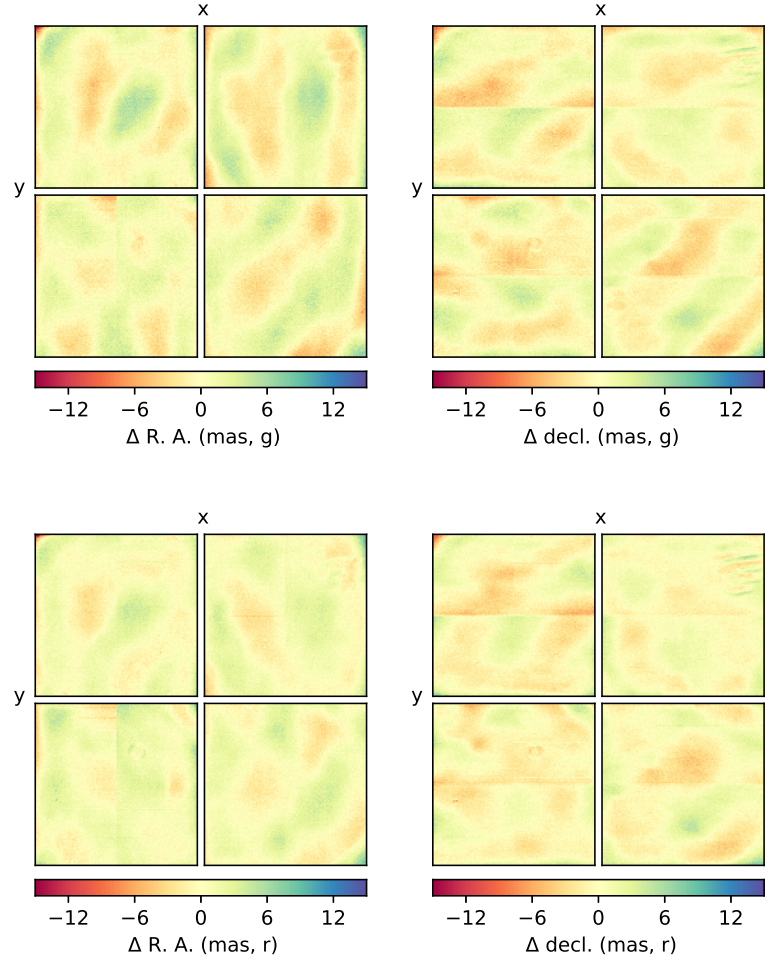}
\caption{The mean residuals across all the $g$ (upper panels) and $r$ band images (bottom panels) are shown. The left panels show R.A. residual maps, and the right panels show decl. residual maps. }
\label{mu15}
\end{figure*}

\begin{figure*}
\centering
\includegraphics[width=10cm]{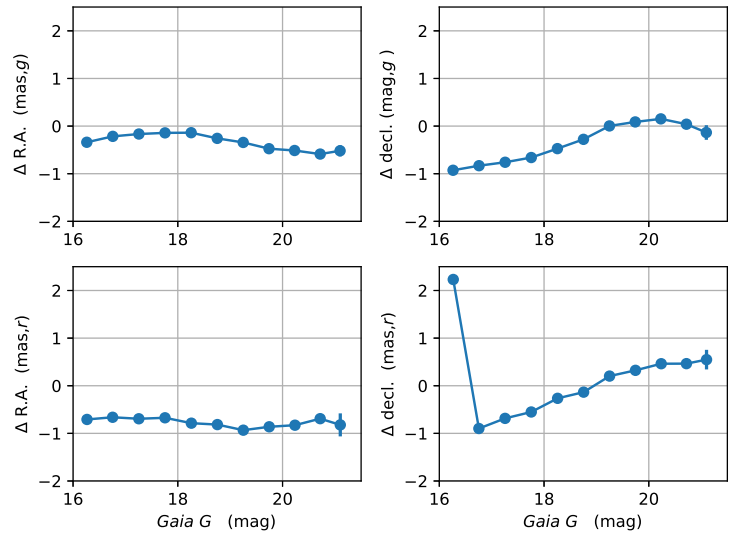}
\caption{The mean and error in the mean of the position residuals vs. \emph{Gaia} magnitude for images taken under $\sim 1.5''$.}
\label{mu16}
\end{figure*}

\begin{figure*}
\centering
\includegraphics[width=10cm]{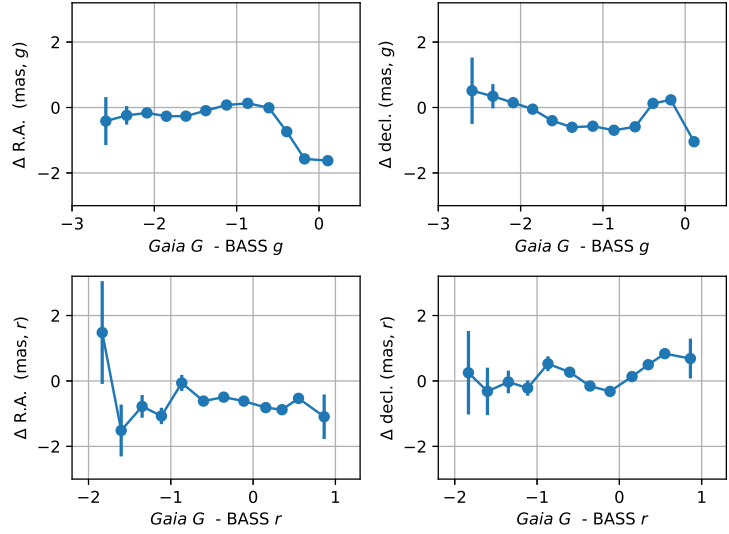}
\caption{The mean and error in the mean of the position residuals vs. color for images taken under $\sim 1.5''$ is shown.}
\label{mu17}
\end{figure*}

\begin{figure*}
\centering
\includegraphics[width=12cm]{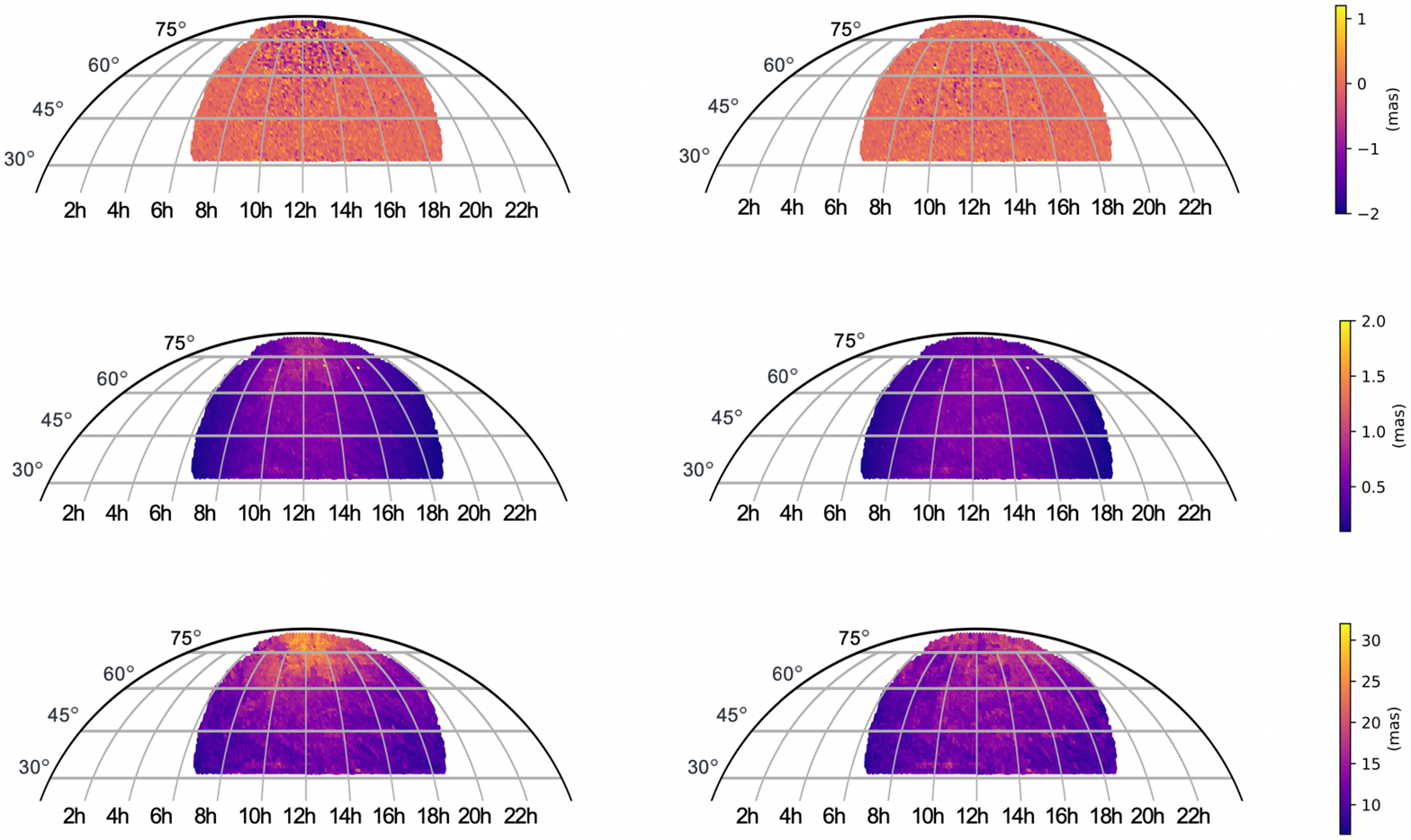}
\caption{The variation of the systematic and random astrometric errors in the position (right panels, R.A.; left panels, decl.) as a function of the equatorial coordinates. Darker colors indicate lower errors. }
\label{mur8}
\end{figure*}

\section{ RESULTS}

\subsection{Positional errors as a function of measured position, magnitude and color}

The relationship between the position error and the measured position was analyzed by the residual map, which was created by measuring the average residual between BASS and \emph{Gaia}. All the BASS data were used to create residual maps for the $g$ and $r$ bands, as shown in Figure 9. The left panels show R.A. residual maps, and the right panels show decl. residual maps. Similar to this figure is Figure 10 from the BASS DR1, which shows a large offset in the middle of the decl. residual map because of the lack of low CTE correction. In contrast to DR1, the shift in the middle of the decl. residual maps in Figure 9 becomes smaller, with the maximum shift decreasing from 60 mas to a few mas, indicating that most of the low CTE induced centroid shift has been removed. We did not apply the residual maps to the final coordinates because they degraded astrometric performance for some images. 

Images taken under similar conditions have similar random errors, making any systematic errors more apparent. Thus, we used images taken from the seeing of $\sim 1.5''$ to test the magnitude and color-dependent position errors. Figure \ref{mu16} shows the mean and error in the mean of the position residuals between BASS and \emph{Gaia} as a function of magnitude. The left columns show the distribution of the mean of the residuals in R.A., and the right columns show the distribution of the mean of the residuals in decl.  For the $g$ band, the systematic errors in R.A. and decl. are small and vary with the magnitude, with a maximum absolute value of about 1 mas. For $r$ bands, the mean R.A. residuals are approximately $-0.8$ mas. The mean of the decl. residuals for stars with magnitudes less than 17 is about 2 mas. One possible reason for this is that $r$ band observation is generally performed under a brighter sky background, and the centroiding of the bright light source may be affected by saturation. The systematic errors for the remaining stars are smaller and vary with magnitude, up to 1 mas. The error in the mean of the position residuals is too small to be shown in the Figure \ref{mu16}.  Figure \ref{mu17} shows the mean and error in the mean of the position residuals between BASS and \emph{Gaia} as a function of color for $g$ and $r$ images, respectively. For both $g$ and $r$ bands, there are small color-dependent systematic errors,  not exceeding 2 mas for R.A. and 1 mas for decl.

\subsection{ Positional errors as a function of Celestial coordinates}

We used the $r$ band data to evaluate the systematic and random errors of the BASS astrometry as a function of celestial coordinates by comparing BASS results with \emph{Gaia}, as shown in Figure \ref{mur8}. We plot the mean systematic offset and the uncertainty of offset in the top and middle panels. The bottom panels show the random errors. Darker colors indicate lower error. As can be seen from the figure, the systematic and random errors  rely on the celestial coordinates. The systematic offset of R.A. in some high declination regions is relatively large, and some of them can reach $-3$ mas. The random errors of R.A. and decl. increase in the area north of declination 60$^{\circ}$. In the region between 30 and 60 degrees of declination, the systematic errors are $-0.01 \pm 0.5$ mas in R.A. and $0\pm0.5$ mas in decl. The random errors are 11.7 mas in R.A. and 11.6 mas in decl. In the region north of declination 60 degrees, the systematic errors are $-0.07\pm0.7$ mas in R.A. and $0\pm0.6$ mas in decl. The random errors are 17.4 mas in R.A. and 14.7 mas in decl. Large systematic and random errors at high declination is likely because of typical observations in these high airmass regions.

\section{SUMMARY}
We describe the astrometric properties of the BASS survey in detail in this paper. From the residual analysis, we find that the BASS suffers from serious optical distortion, low CTE, and DCR. The distortion is unstable during the BASS observation, which relies on the observation time and filter. The low CTE and DCR affect centroiding and are modeled using residuals between the BASS and \emph{Gaia} positions. For the BASS astrometry, the low CTE and DCR were calibrated in measured coordinates, and the calibrated measured positions were then used to calibrate the astrometry against \emph{Gaia}. For each image, we used a fourth-order polynomial to remove optical distortion. After correction for low CTE, DCR and optical distortion, BASS astrometry has little magnitude, color, or position-dependent errors compared to \emph{Gaia}. The systematic error depending on color or magnitude is less than 2 mas. The position error varies with the celestial coordinates and is estimated to be about $-0.01\pm0.7$ mas in the region between 30 and 60 degrees of declination and up to $-0.07$ $\pm$0.9 mas in the region north of declination 60 degrees.  The random error is about 16 mas in the low declination region and up to about 22 mas in the high declination region. The final astrometry results show that the long integration times, coupled with the high BASS stellar density and low CTE, DCR and optical distortion correction, are sufficient to eliminate most instrument and atmospheric effects. 

The BASS is a collaborative program between the National Astronomical Observatories of Chinese Academy of Science and Steward Observatory of the University of Arizona. BASS is a key project of the Telescope Access Program (TAP), which has been funded by the National Astronomical Observatories of China, the Chinese Academy of Sciences (the Strategic Priority Research Program $"$ The Emergence of Cosmological Structures $"$ Grant $\#$ XDB09000000), and the Special Fund for Astronomy from the Ministry of Finance. The BASS is also supported by the External Cooperation Program of the Chinese Academy of Sciences (Grant $\#$ 114A11KYSB20160057) and the Chinese National Natural Science Foundation (Grant $\#$ 11433005). This work is supported by the National Natural Science Foundation of China (NSFC; grant Nos. 11733007, 11673027, 11873053, 12073035,12120101003,12173069, and 11703065), the National Key R$\&$D Program of China No. 2019YFA0405501 and Beijing Municipal Natural Science Foundation, grant No. 1222028. This work is also supported by the Youth Innovation Promotion Association CAS with Certificate Number 2022259, the grants from the Natural Science Foundation of Shanghai through grant 21ZR1474100. We acknowledge the science research grants from the China Manned Space Project with NO.CMS-CSST-2021-A12, NO.CMS-CSST-2021-B10 and NO.CMS-CSST-2021-B04.

\appendix

\section{Field Distortion Correction}

\emph{Distortion correction is an important step in BASS astrometric processing. The specific processing steps are as follows. The geometric distortion correction is performed after CTE and DCR correction using SCAMP. } Equation (\ref{eq1}) and (\ref{eq2})  shows  a linear transformation between corrected measured $x_c$, $y_c$ after low CTE and DCR correction and  intermediate coordinates xi, eta. 
\begin{equation} 
\label{eq1}
\begin{split}
xi \!=\! CD1\_1(x\_c - CRPIX1) + CD1\_2 (y\_c - CRPIX2)
\end{split}
\end{equation}

\begin{equation} \label{eq2}
\begin{split}
eta \!=\! CD2\_1 (x\_c - CRPIX1) + CD2\_2  (y\_c - CRPIX2)
\end{split}
\end{equation}
We find a good description of the distorted tangent plane for the Bok focal plane when we assign each CCD a 4rd order polynomial intermediate longitudinal and latitudinal corrections.  Equation (\ref{eq3}) and (\ref{eq4})  shows high order polynomial distortion correction. The corrections coefficients are keyword PV$_{i\_j}$, where i = 1, 2 are axis indices, and j = 0, . . . ,16  specify indices to the numeric values for the terms in the polynomials. The polynomials are functions of xi, eta. The full equations for the corrections are given as follows:

\begin{equation} \label{eq3}
\begin{split}
\xi' \!=\! PV_{1\_0} + PV _{1\_1}xi + PV_{1\_2}eta +
PV_{1\_4}xi^{2} + PV_{1\_5}xi eta + PV_{1\_6}eta^{2} + PV_{1\_7}xi^{3} +  
PV_{1\_8}xi^{2}eta + \\ PV_{1\_9}xieta^{2} +  PV_{1\_10}eta^{3}  + 
PV_{1\_12}xi^{4} + PV_{1\_13}xi^{3}eta + PV_{1\_14}xi^{2}eta^{2}+
PV_{1\_15}xieta^{3} + PV_{1\_16}eta^{4} 
\end{split}
\end{equation}

\begin{equation} \label{eq4}
\begin{split}
\eta' \!=\! PV_{2\_0} + PV _{2\_1}xi + PV_{2\_2}eta  +
PV_{2\_4}xi^{2} + PV_{2\_5}xieta + PV_{2\_6}eta^{2} + PV_{2\_7}xi^{3} +
PV_{2\_8}xi^{2}eta +\\ PV_{2\_9}xieta^{2} +  PV_{2\_10}eta^{3} + 
PV_{2\_12}xi^{4}  + PV_{2\_13}xi^{3}eta + PV_{2\_14}xi^{2}eta^{2}
PV_{2\_15}xieta^{3} + PV_{2\_16}eta^{4}
\end{split}
\end{equation}
where $\xi'$ and  $\eta'$  is the distortion-corrected  longitudinal and latitudinal intermediate  coordinate.    A tangent plane projection is applied to the intermediate coordinates  $\xi'$, $\eta'$ to get celestial coordinates.

\end{document}